\documentclass[twocolumn]{aastex63}

\usepackage[T1]{fontenc} 
\usepackage{courier}
\usepackage{amsmath}
\usepackage{enumerate}

\usepackage{xcolor} 
\hypersetup{linkcolor=magenta} 

\newcommand{\Fig}{Figure~}
\newcommand{\Tab}{Table~}
\newcommand{\Eq}{Equation~}

\usepackage{dutchcal}

\begin{document}

\shorttitle{Spatial power spectral analysis of the \textit{Suzaku} X-ray background}

\shortauthors{Y. Zhou et al.}

\title{Spatial Power Spectral Analysis of the \textit{Suzaku} X-ray Background}
\author[0000-0002-5793-554X]{
Yu Zhou
}
\affiliation{Institute of Space and Astronautical Science, Japan Aerospace Exploration Agency, 3-1-1 Yoshinodai, Sagamihara, Kanagawa 252-5210}
\author[0000-0002-7107-8468]{
Kazuhisa Mitsuda
}
\affiliation{National Astronomical Observatory of Japan, 2-21-1 Osawa, Mitaka, Tokyo 181-8588}
\author[0000-0003-4885-5537]{
Noriko Y. Yamasaki
}
\affiliation{Institute of Space and Astronautical Science, Japan Aerospace Exploration Agency, 3-1-1 Yoshinodai, Sagamihara, Kanagawa 252-5210}

\correspondingauthor{Yu Zhou}
\email{zhou.yu@jaxa.jp}

\begin{abstract}
Power spectra of spatial fluctuations of X-ray emission may impose constraints on the origins of the emission independent of that from the energy spectra. 
We generated spatial power spectrum densities (PSD) of blank X-ray skies observed with \textit{Suzaku} X-ray observatory utilizing the modified $\Delta$-variance method.  
Using the total measured count rate as the diagnostic tool, we found that
a model consisting of the sum of two components, one for the unresolved faint point sources and one for the uniform flat-field emission,
can well represent the observed PSD in three different energy bands (0.2-0.5 keV, 0.5-2 keV, and 2-10 keV); 
only an upper limit is obtained for the latter component in 2-10 keV. 
X-ray counting rates corresponding to the best-fit PSD model functions and diffuse emission fractions were estimated,  
and we confirmed that the sum of the counting rates of two model components is consistent with those actually observed with the detector for all energy bands.  
The ratio of the flat-field counting rate to the total in 0.5-2 keV, however, is significantly 
larger than the diffuse emission fraction estimated from the model fits of energy spectra. 
We discussed that this discrepancy can be reconciled by systematic effects in the PSD and energy spectrum analyses.
The present study demonstrates that the spatial power spectrum analysis is powerful in constraining the origins of the X-ray emission.

\end{abstract}


\section{INTRODUCTION}
\label{section:1}

The origin of the cosmic X-ray background (CXB) has been meticulously studied since its first
discovery in the early 1960s\citep{Giacconi1962}. With the deep exposure of \textit{ROSAT}, \textit{Chandra} 
and \textit{XMM-Newton}, a substantial fraction, yet not all, of the CXB in soft (0.5-2 keV) and hard (2-10 keV) 
energy bands has been resolved as discrete source emissions\citep{Moretti2003,Lehmer2012}.
There is consensus that the CXB above 2 keV will be eventually resolved into faint sources in the future 
with high spatial resolution, high sensitivity observations and that 
these discrete sources are primarily faint AGNs and soft X-ray galaxies,
which have been confirmed with their multi-wavelength counterparts\citep{Barger2001a,Barger2001b}.
However, the CXB below about 2 keV is considered to contain hot gas emission
from our Galaxy since the first detection in 1968\citep{Bowyer1968, McCammon&Sanders1990}. 
The local emission from the Galaxy is considered to consist of three components: the solar wind induced charge exchange (SWCX) emission from the Heliosphere
\citep{Lisse1996,Cravens2001,Snowden2009,Koutroumpa2011}, 
the thermal emission from the hot gas ($kT \sim 0.1$ keV) in the local bubble extending to about a few 100 pc scales
\citep{McCammon1990,Snowden1998,Snowden1998b,Galeazzi2014,Liu2017}, 
and the emission from the hot gas ($kT \sim 0.2$ keV) beyond the bulk of the galactic neutral interstellar medium
\citep{Pietz1998,Kalberla1998,Kuntz&Snowden2000,Yoshino2009}.

All of the aforementioned local emission components are spatially extended and contain emission lines. 
Some of the strong emission lines have been resolved from the continuum emission with the X-ray microcalorimter 
spectrometer onboard the XQC sounding rocket experiment and also with the X-ray CCD 
(Charge Coupled Device) instruments onboard the \textit{XMM-Newton}, \textit{Chandra}, and \textit{Suzaku} observatories.  
Those spectral features are key to determine the fraction of Galactic components in the CXB.  
However, the results are highly dependent on the emission model, in particular, the metal abundance 
of the hot gas assumed in the analysis and the spectral model fits. 

In addition to the Galactic diffuse emission, the WHIM (Warm-Hot Intergalactic Medium, \citet{Cen1999}) 
may contribute to the CXB below 2 keV.  The WHIM emission is also spatially extended, however, 
the length scales of the spatial variation is likely different from those of the Galactic diffuse components.

A few studies attempted to evaluate the WHIM contribution to the diffuse X-ray background 
using angular clusterings of the WHIM in the unresolved CXB, although no consensus has been reached thus far.
\citet{Galeazzi2009} measured the X-ray background in \textit{XMM-Newton} satellite and reported
that 12\%$\pm$5\% of the diffuse emission in 0.4-0.6 keV energy range is owing to the intergalactic medium. 
Forecasting from the hydrodynamic simulation model, \citet{Ursino2006} estimated the characteristic 
angular size of the WHIM to be less than a few arcminutes by the autocorrelation function. 
\citet{Cappelluti2012} investigated the cosmic X-ray background using 4 Ms deep observations
of the \textit{Chandra Deep Field}-South with the power spectral analysis and 
estimated that $\sim$55 \% of the unresolved CXB flux, i.e. the CXB flux after removing contributions 
of all the point source resolved with the instrument, originates from the 
intergalactic medium (IGM) in 0.5-2 keV band, and a third of which is produced by the WHIM ($10^5<T<10^7$ K and density
constrast $\delta < 1000$). 

In this paper, we present the spatial power spectral analysis of \textit{Suzaku} blank-sky observations.  
Since the spatial resolution of the X-ray telescopes is limited ($\sim 1$ arc min) compared to 
those of \textit{Chandra} and \textit{XMM-Newton} ($\sim 1 - 15$ arc sec, we will not be able to resolve the clustering of the WHIM.  
However, we expect that we can distinguish the emission of the unresolved point sources from the spatially extended Galactic emission, 
and the result is not susceptible to the spectral model that is assumed. 
We describe our dataset and the method in section \ref{section:2}. The results are presented 
and discussed in section \ref{section:3}. We compare our result with ROSAT R4 band in section \ref{section:4}. In section \ref{section:5}, we present the conclusion.

\section{Analysis Methods}
\label{section:2}

\begin{figure*}[ht]
\centering
\includegraphics[width=1\textwidth]{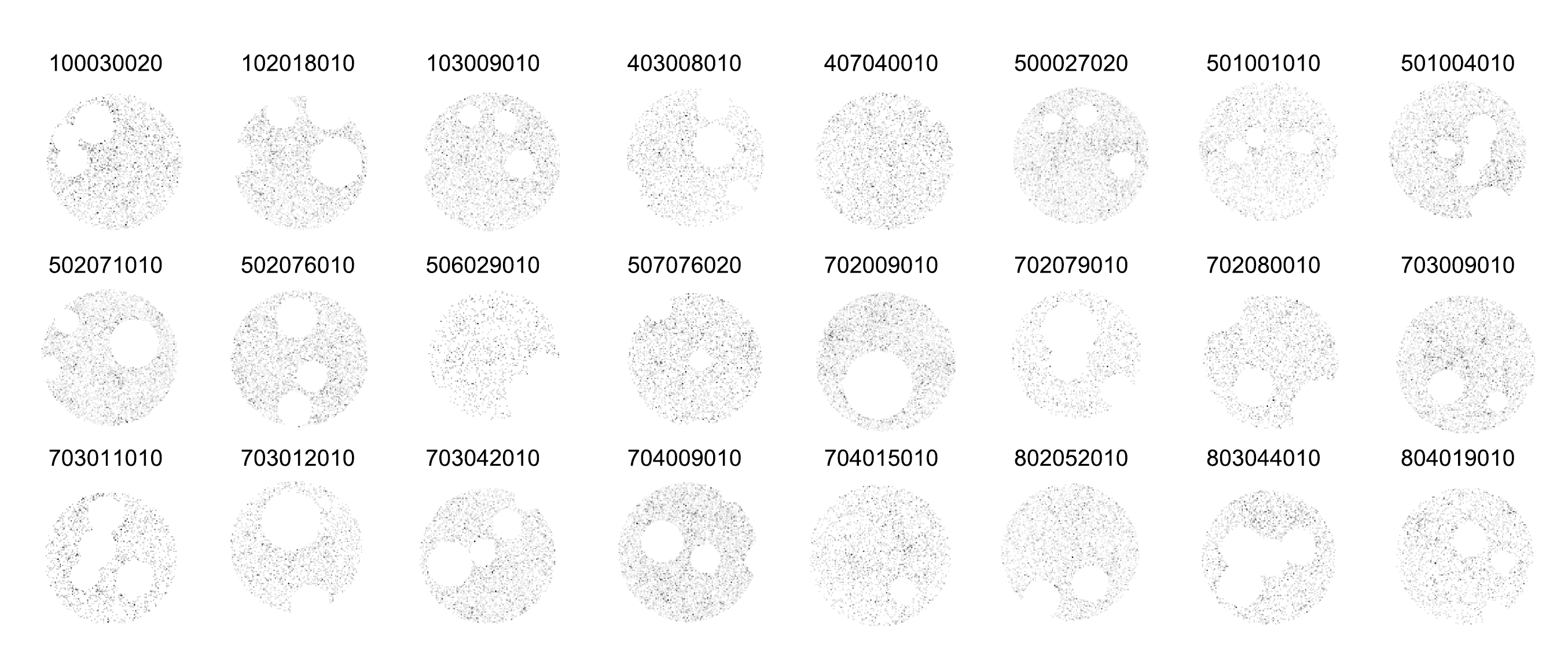}
\caption{Selected \textit{Suzaku}(XIS1) masked images in the 0.5-2.0 keV range. Every extracted
image has its own unique mask for the power spectrum analysis.}
\label{fig:suzaku_sample}
\end{figure*}

\subsection{Data reduction}
\label{section:2.1}
Based on a sample of 187 \textit{Suzaku} X-ray imaging spectrometer (XIS) observations of the diffuse X-ray background 
from 2005 to 2012 \citep{Sekiya2014}, we further screen the CXB data to obtain a robust
mask that excludes any distinguishable source-related gradient in the field. 
First, the images of the \textit{Suzaku}/XIS observations are produced following the standard 
full XIS reprocessing and screening routine via the \ttfamily{Aepipeline} \normalfont command integrated
in \textit{Suzaku} \ttfamily{FTOOLS} \normalfont in \ttfamily{HEASoft}\normalfont(version 6.26.1).
With the latest calibration database(last modified on Oct 10, 2018 for XIS)\footnote{https://heasarc.gsfc.nasa.gov/docs/heasarc/caldb/caldb$\_$sup- ported\_missions.html}
any illuminated corners caused by calibration sources or hot and flickering pixels have been fully removed
in the clean event files. These images are then sent to the manual mask selection to ensure that: 
(1) any resolved point source and vicinity region will be excluded with the masks;
(2) any suspicious spots that are systematically brighter than the average fluctuation field will be removed using masks;
(3) any observation that has shown a clear large gradient over the image, instead of a background fluctuation, will be discarded;
(4) any observation that has less than a 30\% effective area left after being masked will be discarded.
Later on, we performed a visual inspection of the blank sky sample and no significant features are seen in the masked images.
We show some examples of masked images of \textit{Suzaku} observations in \Fig \ref{fig:suzaku_sample}.

For most \textit{Suzaku} observations, images from the detector XIS0, XIS1, and XIS3 are available,
given that XIS2 has not been functioning since 2006. We discarded all XIS2 images and 103 
observations after the screening process, remaining 84 images for XIS0, 1, 3, respectively, for the power spectrum calculation,
of which the accumulated observation time equals to $\sim$1 Ms for each detector. As a thinned backside-illuminated (BI) 
device, XIS1 has a higher sensitivity to X-ray photons at soft energy bands compared with the front-side illuminated (FI) chips,
XIS0 and XIS3. We analyzed the images from different detectors separately.

\subsection{Method to estimate the spatial power spectrum density}
\label{sec:PSD_method}
To robustly evaluate the fluctuation of the diffuse background at various scales for the images with 
gaps and holes, we adopt a modified $\Delta$-variance method to calculate
the power spectrum of the image using a mask \citep{Arevalo2012}. The masked image is
convolved with a Mexican hat filter, which is equivalent to the difference between two Gaussian functions
with smoothing lengths $\sigma_1 = \sigma / \sqrt{1+\epsilon}$, $\sigma_2 = \sigma \sqrt{1+\epsilon}$,
where $\epsilon \ll 1$. The filter can be described by  
\begin{eqnarray}
\hat{F}_{k_r}(k) & =& G_{\sigma_1}(k) - G_{\sigma_2}(k) = e^{-2\pi^{2} k^2 \sigma_1^2} - e^{-2\pi^{2} k^2 \sigma_2^2}. \label{eq:1} \\
& \simeq & 4 \pi^2 \sigma^2 \epsilon k^2 e^{-2 \pi^2 \sigma^2 k^2} \label{eq:1'}
\label{eq:mexican_hat_limit} 
\end{eqnarray} 
where $k$ is the spatial frequency/wavenumber (the angular length scale $\theta = 1/k$). The shape of the filter frequency response does not depend
on $\epsilon$ in the limit of $\epsilon \to 0$ as shown in \Eq (\ref{eq:mexican_hat_limit}). The filter takes a 
maximum at $k=k_r \equiv 1 / (\sqrt{2} \pi \sigma) = 0.225 / \sigma$, and the response is rather
broad (full width is $\sim k_r$). We adopt $\epsilon = 0.01$ for this study.

The mask, $M$, is defined to be one for the image pixels used for analysis, and to be zero
for the pixels discarded. Convolving the masked image with a Gaussian function will produce 
spurious features. Their amplitudes can be corrected by dividing by the mask that is 
convolved with the same Gaussian function. Therefore, the difference between two corrected images 
will be a good estimation for the fluctuation of the original image at a frequency scale of
$k_r$, thus at a spatial length scale of $\sigma$. 
The variance of the image at that frequency can be estimated as:
\begin{eqnarray}
\centering
V_{k_r,obs} = \frac{N}{N_{(M=1)}} \int (\frac{G_{\sigma_1}*I}{G_{\sigma_1}*M} - \frac{G_{\sigma_2}*I}{G_{\sigma_2}*M})^2 M^2 d^2x,
\label{eq:Variance}
\end{eqnarray}
where $N$ and $N(M=1)$ are total number of original pixels and the adopted
pixels, respectively, and $I$ denotes the masked image in the unit of counts. The symbol *
represents a convolution, namely,
\begin{eqnarray}
G_\sigma *I = \int  G_\sigma(y) I(x-y) d^2y , 
\end{eqnarray}
where $x$ and $y$ are 2-dimensional vectors.

The power spectrum density (PSD) can be estimated by normalizing $V_{k_r,obs}$ by the 
2-dimensional frequency band widths of the filter. Thus, 
\begin{eqnarray}
\centering
P(k_r) = \frac{V_{k_r,obs}}{\int |\hat{F}_{k_r}(k)|^2 d^n k} = \frac{V_{k_r,obs}}{\epsilon^2 \pi k_r^2 }.
\end{eqnarray}
We further renormalized P($k$) so that it represents the PSD of the surface brightness:
\begin{eqnarray}
P_2(k) = \frac{P(k)}{S T^2}
\end{eqnarray}
where $S$ and $T$ are, respectively, the solid angle of the sky selected by the mask(where M=1) and the
exposure time of the observation. In this equation and hereafter, we replace $k_r$ with $k$
for convenience. The characteristic angular length scales range from 37 to 930 arcsecond
to sample the PSD of the \textit{Suzaku} observations, that are limited by the image pixel size and the field of view of the telescope, respectively.
The sampling rate is chosen such that at each frequency the Mexican-hat filter
function of the next point decays to $1/\sqrt{2}$ of its maximum value.

\begin{figure}[t]
\centering
\includegraphics[width=0.5\textwidth]{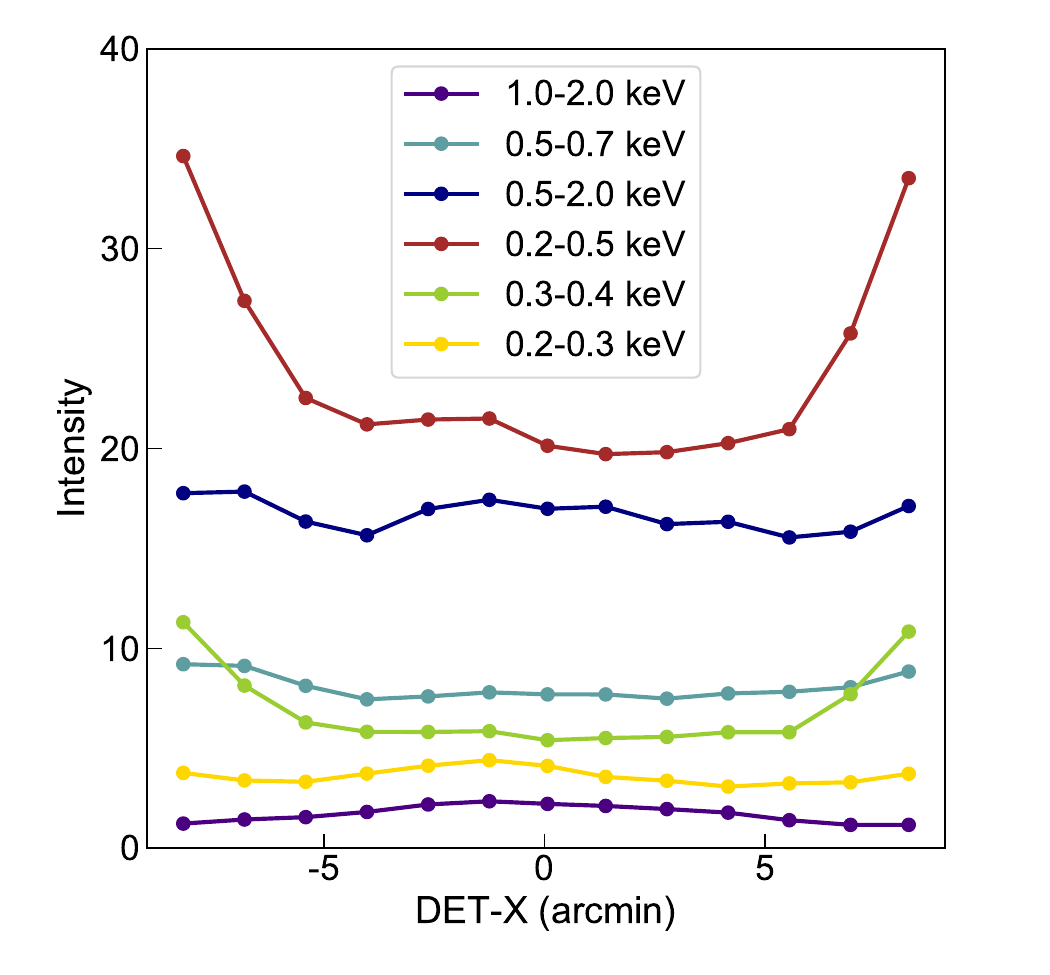}
\caption{Exposure maps for different energy bands. The day-Earth images are accumulated to obtain
the exposure map of the flat input field. The curvature shows the smoothed counts along the
cross section of the middle detector plane.}
\label{fig:exposure map}
\end{figure}

\subsection{Non-X-ray Background subtraction}
The non-X-ray background (NXB) originates from cosmic rays and solar protons in the space hitting 
through the spacecraft, interacting with materials inside, and depositing energy on the CCD detectors.
Owing to the low orbit altitude of the \textit{Suzaku} satellite, the NXB level of \textit{Suzaku}
is significantly less influenced by solar flares and is more stable than \textit{Chandra} and \textit{XMM-Newton}\citep{Yamaguchi2006}.
Moreover, the predicted counts of the NXB for each observation can be reliably estimated based on
the satellite orbital period, given that the NXB is anti-correlated with the cut-off rigidity
and correlated with the count rate of the PIN upper discriminator onboard \textit{Suzaku}\citep{Tawa2008}. 

The spectra and image of the NXB are available from the observation of night-Earth events
for Earth elevation angles less than $\rm -5^o$\citep{Tawa2008}. 
From the database we can construct an NXB event list suited for a particular observation of each of four XIS sensors separately.
The event lists contain night-Earth events within $\pm$ half a year of the observation and 
with a same cut-off rigidity time distribution as the observation. From the events we can
construct NXB images, subtract the NXB images from the observational images, and then 
calculated the spatial PSD for the subtracted images. Alternatively, we may first calculate
the PSDs of the observations and NXB images respectively and calculate the difference.
We tried two methods and found that the difference is negligibly small between the two PSDs. However,
we decided to adopt the first method for the following two reasons. Firstly, the PSD of the
raw image may contain the cross terms of the observed and NXB images, which can not be subtracted
from by NXB PSD subtraction. Secondly, the NXB PSD consists of the Poisson noise component and
the spatial gradient component. The spatial gradient component can be subtracted by the first
method using the NXB image with a much longer accumulation time than the observation, thus the
Poisson noise itself is negligible for those NXB images constructed for spatial gradient PSD subtraction. 
And the Poisson-noise PSD of the NXB in the observations can be subtracted independently by
A-B technique(see next subsection). $I$ in \Eq (\ref{eq:Variance}) is now given by
\begin{eqnarray}
I(x) = I_{\rm obs}(x) - \frac{T_{\rm obs}}{T_{\rm NXB}}I_{\rm NXB}(x)
\end{eqnarray}
where $I_{\rm obs}$ and $I_{\rm NXB}$ are, respectively, the masked-observation and NXB images, and 
$T_{\rm obs}$ and $T_{\rm NXB}$ are their exposure times.

\subsection{Poisson-noise subtraction}
\label{sec:PSD_Poisson}
Spatial fluctuation due to the Poisson counting statistics are significant for both the 
observation and the NXB images, in particular, at high frequencies. We estimated the PSD of 
the Poisson fluctuation using the A-B technique\citep{Kashlinsky2005} and subtracted it from 
the PSD. We divide the observation, which contains the NXB events, into even and odd events in time sequence,
construct even and odd images (A and B images) for each observation, calculate PSD for the (A-B)/2 images,
and subtract them from the PSDs of the NXB-subtracted observation images.

\subsection{Ensemble average and standard deviation of PSD}
\label{sec:ave_sigma_PSD}
Our data consist of in total 252 images of \textit{Suzaku} observations from three XIS(0,1,3) CCD detectors. We average 
all the Poisson-noise-subtracted PSDs for each corresponding XIS chip independently as
\begin{eqnarray}
\bar{P}_2(k) = \frac{1}{n}\sum_{i=1}^n P_{2,i}(k),
\end{eqnarray}
where $P_{2,i}(k)$ is the Poisson noise subtracted PSD of the $i$-th observation, and $n$ is
the total number of PSDs averaged. Then we estimate the standard deviation of the PSD from
the variance of $P_{2,i}(k)$,
\begin{eqnarray}
\sigma_{P_2}(k) = \sqrt{
\sum_{i=1}^{n}
\frac{  \left( P_{2,i}(k)-\bar{P}_2(k) \right)^2 }{n(n-1)}} .
\end{eqnarray}
In the next section, we will perform $\chi^2$ model fitting to $\bar{P}_2(k)$ using $\sigma_{P_2}(k)$
as the standard deviation. Strictly, it is not correct because the NXB events used to estimate 
spatial power which is later subtracted in $P_{2,i}(k)$ partly overlaps with each other. 
However, since the contribution of the NXB spatial gradient to the power spectrum is at most 10\% beyond 150 arc second,
we still use $\sigma_{P_2}(k)$ as the error of $\chi^2$ fitting.

\subsection{Non-uniformity of effective area}
\label{sec:vignetting}
The effective area of the observed images are not spatially uniform because of two reasons:
the vignetting of the telescope\citep{Serlemitsos2007} and the non-uniform thickness of the
contaminants sticked on the optical blocking filter of the XIS detectors, which create extra
absorption in soft X-ray bands. The non-uniformity works as a window function of the Fourier
analysis. Thus the true Fourier transform of the sky image will be convolved with the Fourier
transform of the non-uniformity function, so that the obtained PSD is smoothed by this transfer
function. Figure \ref{fig:exposure map} shows the exposure maps for the flat field in different
energy ranges, illustrating the spatial non-uniformity caused by vignetting effect and contamination.
Given the small gradient in the exposure map image, the frequency band width of this window 
function filter due to the non-uniformity is in general narrower than that of the Mexican-hat filter.
Both filters smooth the PSD but have different influences on the power at zero frequency, since  
the non-uniformity of effective area convolves the PSD at zero frequency, but Mexican hat filter 
only maximize the power at its own smoothing length scale, as shown in \Eq \ref{eq:mexican_hat_limit}.  
The power at zero frequency spreads out to the PSDs at non-zero frequencies by the non-uniformity of effective area
first, and then gets further smoothed with the Mexican-hat filter. 
We took these effects into account when modeling and fitting for the observed PSD (see subsection \ref{section:2.8}).

In order to create model PSD functions, we first constructed simulation event list using
the ray-tracing Monte-Carlo simulator \ttfamily{xissim}\normalfont\ in which both the vignetting
and contamination are considered as functions of photon energy\citep{Ishisaki2007}. In the 
present analysis we extend our energy range down to 0.2 keV, which is not supported in the 
standard analysis package. To extend the energy range we adopted the latest update of the 
contamination calibration. The contamination source is considered to be the outgassed 
plastic material from the anti-vibration rubber protecting the inertial reference unit 
during the launch of the spacecraft. 
The effect of the contamination on the \textit{Suzaku} X-ray spectra below 0.5 keV was extensively 
studied recently by Nagayoshi (2019) (PhD thesis, University of Tokyo), and the calibration data base 
was revised. According to this study, the contaminant contains S in addition to the elements, C, H, and O 
which were previously considered. This study successfully reproduced the 15 X-ray spectra of the 
calibration source, RX J1856.5-3754, in the 0.2-1 keV band observed from 2005 to 2014. 
It was also shown that the new calibration database better reproduces an improved energy spectrum 
of the other calibration source, 1E0102.2-7219, which is subject to a lager Galactic absorption 
than RX J1856.5-3754 in the range of 0.3 to 3 keV. We used the new calibration database in
the ray-tracing simulation.

\subsection{Point spread function of the telescope}
The sky image is convolved with the point spread function (PSF) of the telescope before being
detected by the X-ray CCD cameras. Consequently the Fourier transform of the sky image is
multiplied with the Fourier transform of the PSF and consequently attenuated significantly in the high
frequency range. We also take this effect into account in the PSD modeling by creating the
power spectrum for the point source using the \ttfamily{xissim}\normalfont\ ray-tracing simulation.
The shape of PSF varies across the field of view of the XIS detectors, and the response
is averaged over the detector plane given the wide and random distribution of X-ray sources in the field of view.

\subsection{Model PSDs for unresolved point sources and a uniform flat field}
\label{section:2.8}
We constructed two model functions, one for the unresolved point sources (UPS) and the other for the uniform flat-field
emission (UFF) using Monte-Carlo simulations. The latter model is used to represent the diffuse X-ray background.

To construct a realistic UPS model, we need a spatially-random point source sample whose
brightness distribution obeys the $\log N$-$\log S$ relation of the sky.
Because we average the PSDs from different \textit{Suzaku} observations with different exposure times, 
the detection threshold of the point sources varies from field to field.   
We, however, represent the detection threshold with a single model parameter, $S_{\rm max}$, 
the flux of the point source in corresponding energy band.  
We adopt the $\log N$-$\log S$ relation from \citet{Moretti2003}. We generate simulation point sources whose energy flux 
is between $S_{\rm min}$ to $S_{\rm max}$ in a circular sky region of a 20' diameter.  
The value of $S_{\rm min}$ was $1\times 10^{-17} {\rm ~erg~s^{-1}~cm^{-2}}$ and $1\times 10^{-16} {\rm ~erg~s^{-1}~cm^{-2}}$
for 0.5-2 keV band and for 2-10 keV band, respectively.  We do not have an observed $\log N$-$\log S$ for 0.2-0.5 keV band.
We thus scaled the point source energy flux in the 0.5-2 keV band to that of the 0.2-0.5 keV band
to obtain a reference $\log N$-$\log S$ for this band, assuming an energy spectrum
of an absorbed power-law spectral shape, modeled by \ttfamily{tbabs*po}\normalfont\ in \ttfamily{XSPEC}\normalfont\ with
the hydrogen column density $N_{\rm H}=2.5\times 10^{-20}\rm\ cm^{-2}$ and the photon index $\Gamma=2$.
The flux lower limit $S_{\rm min}$ is chosen to be $1\times 10^{-17} {\rm ~erg~s^{-1}~cm^{-2}}$ for the 0.2-0.5 keV band,
below which the detector is no longer sensitive to receive any photons given the simulation time.
The X-ray emission of the point sources in the chosen field of view is generated
via the Monte-Carlo simulator \ttfamily{xissim}\normalfont. 

To generate a photon list in the simulation, we need to specify an energy spectrum for each
source. We adopted a power law function absorbed by the galactic cold interstellar medium.
The photon index is 2.0 for the sources simulated in all energy bands, and
we assumed a galactic HI column density of $N_{\rm H} = 2.5 \times 10^{20}\rm\ cm^{-2}$ for all sources.

We generated a photon list for 66.5 ks of observation (the average exposure time of the
\textit{Suzaku} samples) and calculated the PSD according to the method described in sections \ref{sec:PSD_method} and \ref{sec:PSD_Poisson}.  
The photon lists for sources over a sky area of 3 $\rm deg^2$ are generated assuming the $\log N$-$\log S$
relation in each energy band. The total numbers of the photon lists are 21085 for $S_{\rm max}=10^{-13} \rm ~erg~s^{-1}~cm^{-2}$ 
in 0.2-0.5 keV, 38289 for $S_{\rm max}=10^{-13.82} \rm ~erg~s^{-1}~cm^{-2}$ in 0.5-2 keV, and
23375 for $S_{\rm max}=10^{-12.99} \rm ~erg~s^{-1}~cm^{-2}$ in 2-10 keV. We selected 14 subsets of
the photon list in a circular sky region of a 20' diameter to mimic different pointing of the observations
and calculated the ensemble average and standard deviation of PSDs as described in section \ref{sec:ave_sigma_PSD}.

\begin{figure}[t]
\centering
\includegraphics[width=0.48\textwidth]{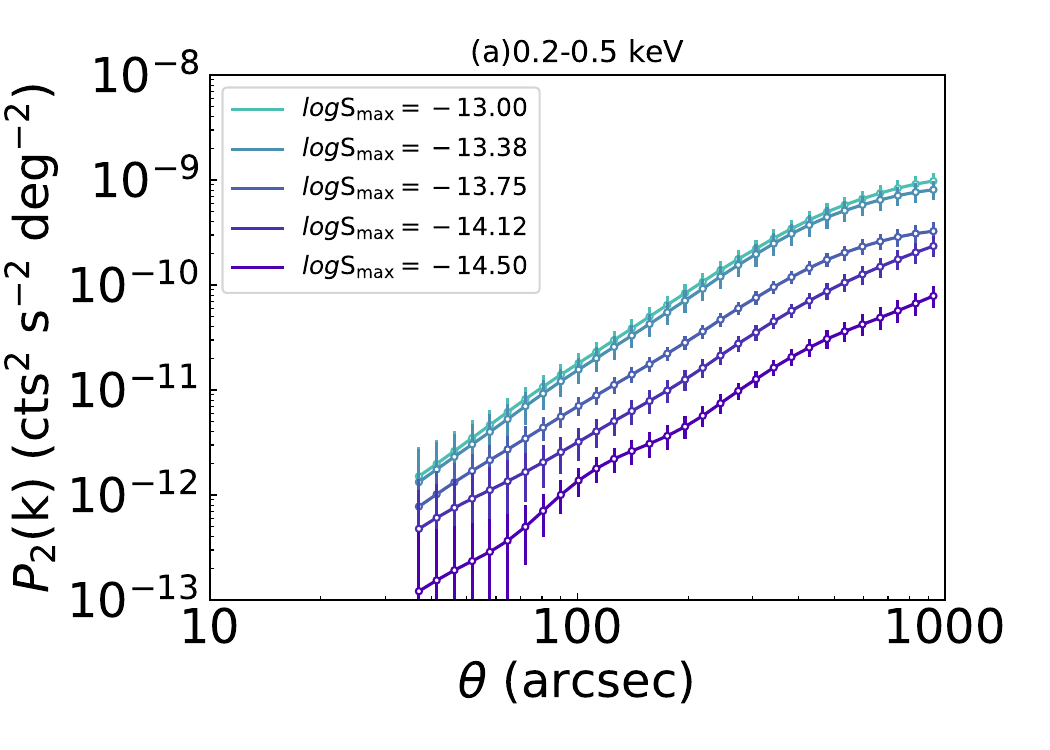}
\includegraphics[width=0.48\textwidth]{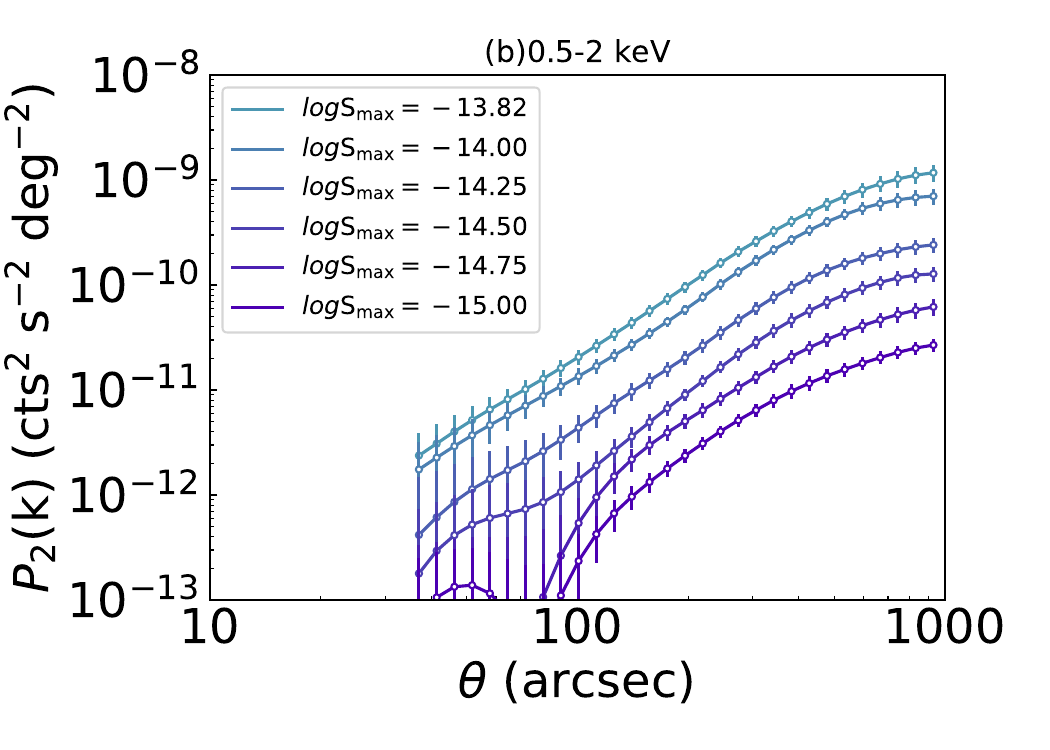}
\includegraphics[width=0.48\textwidth]{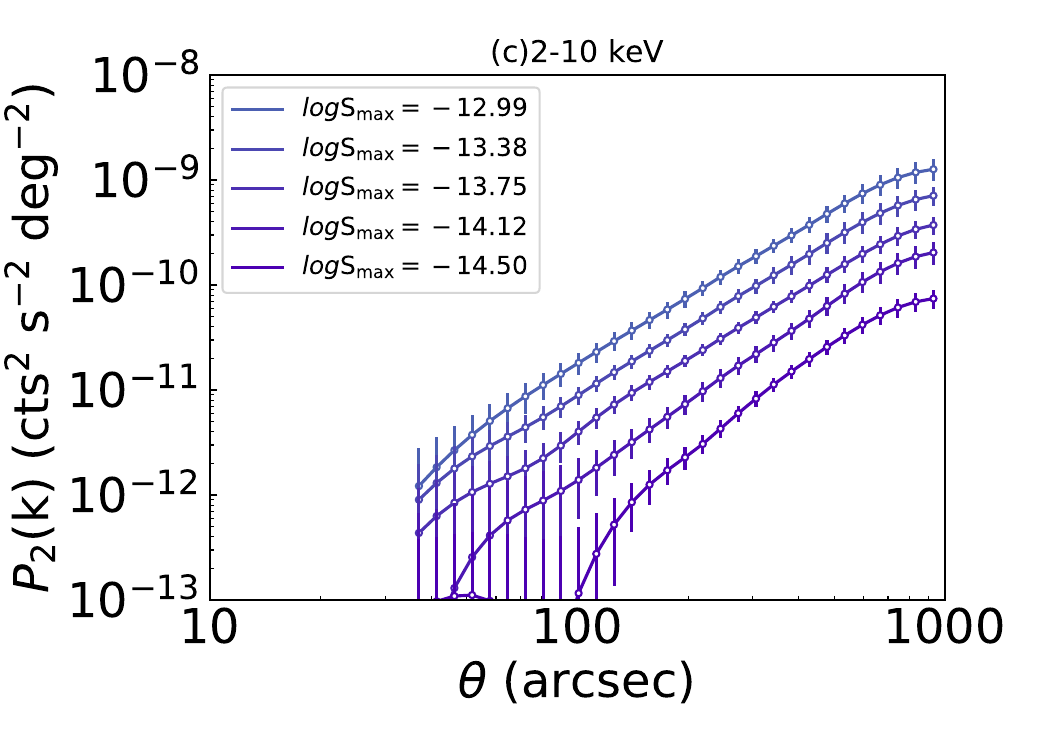}
\caption{PSDs of simulation images of the point sources in three different energy bands (a: 0.2-0.5 keV, b: 0.5-2 keV, c: 2-10 keV). 
The point sources were generated so that their number density follow the $\log N$-$\log S$ relation with an upper boundary of the energy flux, $S_{\rm max}$.  
The amplitude of the PSD is dependent on $S_{\rm max}$ while the shape of PSD as a function of the angular length scale is not. }
\label{fig:logNlogS_point_source_model}
\end{figure}

To fit for the observed PSD spectrum, we generated the average PSD models with a series values of $S_{\rm max}$.
Figure \ref{fig:logNlogS_point_source_model} shows the PSD model functions simulated with various $S_{\rm max}$ 
for 0.2-0.5 keV, 0.5-2 keV, and 2-10 keV energy band.
The free parameter of the UPS model is $S_{\rm max}$ during the fitting procedure. The fitted PSD $S_{\rm max}$
is interpolated linearly from the $S_{\rm max}$ of the PSD models based on the model and the observed PSD spectrum amplitude.

We assumed a simple power-law energy spectrum for all point sources in the simulations.  In reality the spectral shapes are different for different point sources and as a result the average spectrum may be described with the double broken power-law functions as described in section
\ref{sec:energy_spec}.  In order to check the dependence of the analysis results on the assumed spectral shapes of point sources, 
we also created UPS PSD assuming the double broken power-law functions as 
the energy spectrum with model parameters determined in the section.  
We found the difference of all the results described below are well within 
the statistical errors. Thus we will only adopt and discuss the results utilizing the 
simple power-law model for the energy spectrum hereafter.  

We also constructed a model UFF PSD function by the \ttfamily{xissim}\normalfont\ 
ray-tracing simulator, assuming a power-law function with a photon index $\Gamma = 2.0$ without the Galactic absorption for the
energy spectrum.  The size of the flat field is 20 arc minutes as the default setting of \ttfamily{xissim}\normalfont\   
and the XIS was pointed to the center of the flat field in the simulation. The power-law energy spectrum
assumption might be too simple for the diffuse emission, but it is confirmed that the UFF model
count rate and PSD do not depend on the energy spectrum of the input source. That is, to check the energy dependence,
we assumed an alternative spectral model consisting of the sum of an absorbed and an unabsorbed thermal emission as described in section \ref{sec:energy_spec},
and modelled the PSD with the spectral parameters fixed to the best-fit values for \textit{Suzaku} observations.
Then we compared the PSD amplitude and count rate with the ones simulated assuming the simple power-law function. 
As a result, we found that the two UFF PSD models with different energy spectra are identical to each other.  
We thus adopt the PSD model based on the simple power-law spectrum in the following analysis.

Uniform flat-field (UFF) emission intrinsically contains power only at zero frequency.  However, we see non-zero power at non-zero frequencies owing to the non-uniformity of the effective area (see section \ref{sec:vignetting}). 
The surface brightness of the UFF model can be arbitrary since the PSD amplitude of the UFF model 
is proportional to the square of the model surface brightness, or the count rate. 
The uniform diffuse emission component in the observed PSD will be represented by the UFF model
that is scaled with a normalization factor, which is the actual free parameter of the PSD fitting.

\section{Results}
\label{section:3}

\subsection{Sky averaged PSD in three energy bands}
\label{section:3.1}

\begin{figure*}[ht]
\centering
\includegraphics[width=1\textwidth]{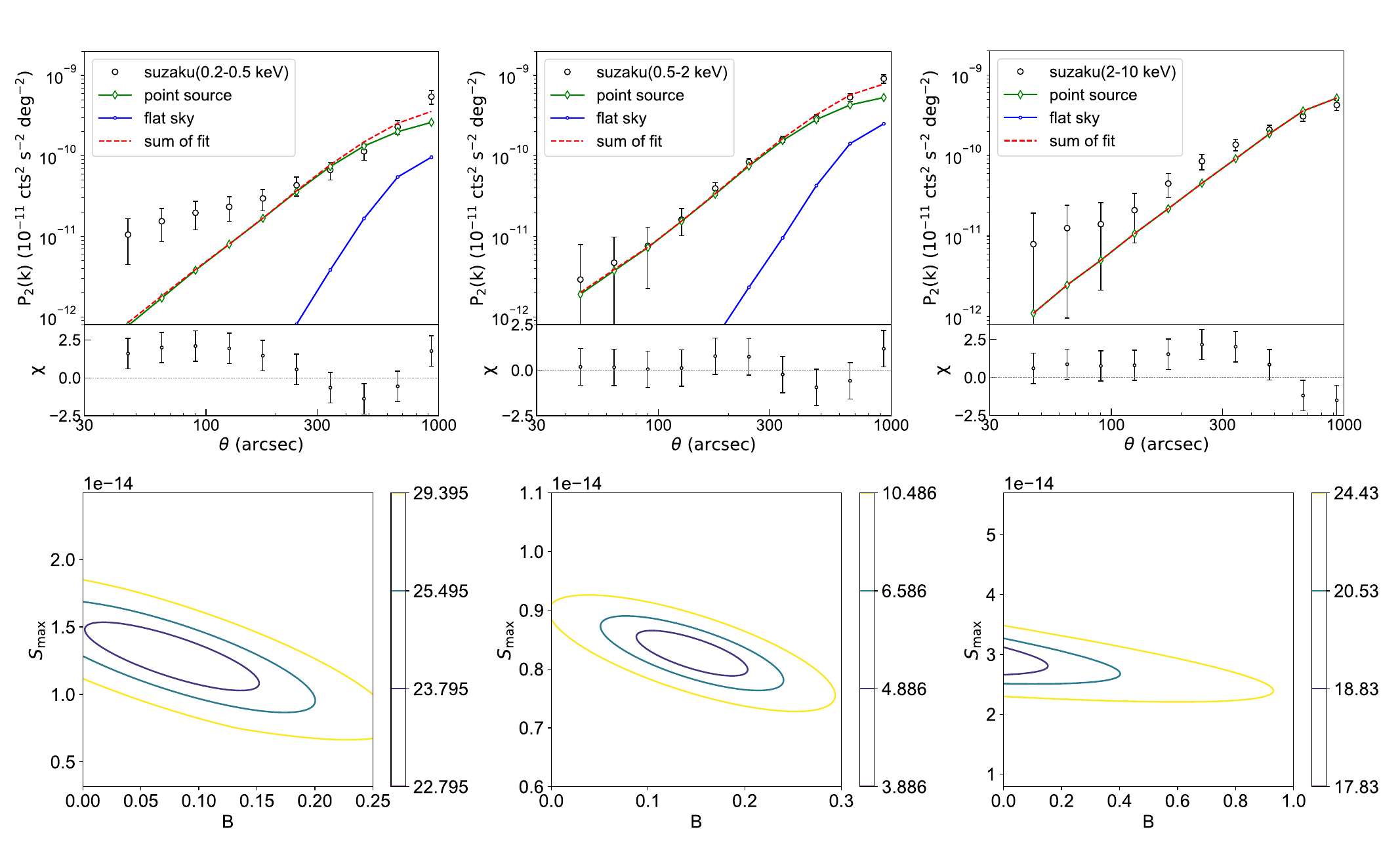}
\caption{Top panels: Power spectra of the \textit{Suzaku} observation in three energy bands ($left$:0.2-0.5 keV, $middle$:0.5-2 keV, $right$:2-10 keV)
with the best-fit PSD model and each components. The residuals of the fits, (data - model)/error, are also shown.
The UPS model component was created by ray-tracing simulation assuming a random distribution of the unresolved faint point sources which follow the $\log N - \log S$ relation and the flux cut off, $S_{\rm max}$.   
The UFF model is simulated assuming a uniform flat field as the input.
Bottom panels: $\chi^2$ contour maps of the fits.  The three contours represents single parameter errors at 1-$\sigma$, 
90\% and 99\% confidence limits, i.e. $\chi^2_{\rm min} + 1, +2.7$, and $+6.6$, respectively.}
\label{fig:ps-fit-chisq}
\end{figure*}

The spatial PSDs averaged over 252 observation field were obtained in three different energy
bands for XIS1 (back-illuminated CCD) and in two bands for XIS0 and XIS3. The PSDs of three
different detectors are consistent with each other after correcting for the difference in
the effective area. 
We concentrate on the XIS1 results to study the energy dependence of the PSD.
Poisson noise subtracted PSD are shown in Figure \ref{fig:ps-fit-chisq} 
with 1-$\sigma$ error bars, where the x-axis of the plot is the length scale $\theta$ (the quantity
denoted with $k$ in section \ref{sec:PSD_method}).  

\subsection{PSD model fitting}
\label{section:3.2}

In order to find out whether the observed PSD can be explained by a sum of spatial fluctuations due to the unresolved point sources and due to the flat field, 
we fitted them with the model function:
\begin{eqnarray}
PSD(\theta) = PSD_{\rm UPS}(\theta, S_{\rm max}) + {\rm B}\ PSD_{\rm UFF}(\theta) ,
\end{eqnarray}
where $PSD_{\rm UPS}$ and $PSD_{\rm UFF}$ are respectively the $PSD$ models for UPS and UFF of the
relevant energy band, and both are functions of the angular length scale, $\theta$.  
$S_{\rm max}$ and $B$ are free parameters of the fit. 
We minimized the $\chi^2$ and obtained the results shown in Table \ref{table:PSD_fit}.

\begin{deluxetable}{cccc}[h] %
\tablenum{1}
\tablewidth{0pc}
\tabletypesize{\scriptsize}
\label{table:PSD_fit}
\tablecaption{PSD fitting result}
\tablehead{\colhead{Energy band (keV)} & \colhead{0.2-0.5} & \colhead{0.5-2} & \colhead{2-10}}

\startdata
$S_{\rm max}^{\dagger}$  & ﻿$﻿1.2_{-0.2}^{+0.2}$ & ﻿$﻿﻿0.82_{-0.04}^{+0.04}$  & ﻿$﻿2.9_{-0.2}^{+0.2} $  \\
B  & ﻿$﻿0.077_{-0.075}^{+0.074}$ & ﻿$﻿0.14_{-0.05}^{+0.05}$  & ﻿$﻿0^{+0.15} $  \\
$\chi^2$  & ﻿$22.8$ & ﻿$﻿﻿3.88$  & ﻿$﻿﻿17.8$ \\
d.o.f  & ﻿$8$ & ﻿$8$  & ﻿$8$ 
\enddata
\tablenotetext{\dagger}{In the unit of ${\rm 10^{-14}~ erg~ s^{-1}~ cm^{-2}}$ }
\tablenotetext{}{Quoted errors are single-parameter 1-$\sigma$ errors. }
\end{deluxetable}

The best fit model functions and the residuals are shown in Figure \ref{fig:ps-fit-chisq}. 
The model function represents the observed PSD generally well. 
The minimum $\chi^2$ values for 0.2-0.5 and 2-10 keV are large: the upper tail probabilities are respectively 0.36\% and 2.3\%.
We see in \Fig\ref{fig:ps-fit-chisq} a systematic deviation of the data points over the model at the length scales below $^<_\sim 200''$ are affecting the fit results. 
This deviation extends down to the shortest length-scale bin, thus the length scale of the fluctuation is shorter than that of the PSF of the telescope.  
The large error bars of these bins suggest a large field-to-field variations. 
We thus consider that some flickering pixels of the X-ray CCD camera that could not be removed 
in 0.2-0.5 keV band is the reason to cause the deviation on the short length-scale.
Flickering pixels are the CCD pixels that intermittently generate output pulse heights 
higher than the hot-pixel threshold even without input signals, which are not usable for observations
and have to be disregarded in scientific analysis. The hot-pixel rejection is processed well
above 0.5 keV but might not be as clean for 0.2-0.5 keV since the calibration is only
lately updated for this energy band.

The best-fit values of $S_{\rm max}$ seem to be reasonable when we compare the energy flux distribution of point sources detected in the observation and removed from the PSD analysis.  
The cut off of the flux distribution is at
around $5\times 10^{-15}$, $2\times 10^{-14}$, and $6\times 10^{-14} {\rm~ erg~ s^{-1}~ cm^{-2}}$ for 0.2-0.5 keV, 0.5-2 keV, and 2-10 keV band, respectively.

We can estimate the counting rates of the best-fit PSD models of respective energy bands from the photon lists created by \ttfamily{xissim}\normalfont.  
For UPS model, this can be readily done by interpolating the counting rates from simulations of nearby $S_{\rm max}$ values.
For the UFF model, the counting rate of the model component scales with $\propto \sqrt{B}$.
We thus obtain the counting rates of both UPS and UFF emission that produces the observed PSD.  The total counting rate of each energy band is just the sum of the counting rates of two components. 
The 1-$\sigma$ error intervals for the fitting parameters, $S_{\rm max}$ and $B$, are determined using the $\chi^2$ contour map shown in \Fig \ref{fig:ps-fit-chisq}, 
where the minimum and maximum values of $S_{\rm max}$ and $B$ are picked among all their combinations satisfying $\chi^2 \leq \chi^2_{\rm min}+1$.
We also estimated the counting rate from the energy spectra of the same observations (see subsection \ref{sec:energy_spec})
and show the results in \Tab\ref{tab:count_rates}.  
We find that the total counting rates from the PSD fitting and the energy spectrum are consistent with each other within the 1-$\sigma$ statistical errors.   
According to the PSD fit results in the table, we further estimate the fractions of the flat-field diffuse emission to the total for the three energy bands,  
which are $60_{-42}^{+9}$\%, $56_{-7}^{+5}$\%, and $ <23$\% in 0.2-0.5, 0.5-2, and 2-10 keV, respectively. The 1-$\sigma$ error intervals are again estimated by the $\chi^2$ contours in \Fig \ref{fig:ps-fit-chisq}.   

We obtained only an upper limit for the diffuse emission in 2-10 keV bands, which means that the PSD can be explained with merely unresolved faint point sources.  
For 2-10 keV band, this is consistent with the present understanding of the Cosmic X-ray
Background above 2 keV \citep{Moretti2003,Lehmer2012,Gilli2007}.
For 0.2-0.5 keV band, the uncertainty of the fitted diffuse fraction is large.  However, at a 1-$\sigma$ confidence limit, at least 18\% diffuse emission is supposed to exist in this band.
For 0.5-2 keV band, the PSD cannot be explained with only the unresolved point sources and 
an additional component representing a flat diffuse emission is needed.

\begin{deluxetable}{lccc}[h] %
\tablenum{2}
\tablewidth{0pc}
\tabletypesize{\scriptsize}
\label{tab:count_rates}
\tablecaption{Counting rate estimation from the PSD and the energy spectral fits}
\tablehead{\colhead{Energy band (keV)} & \colhead{0.2-0.5} & \colhead{0.5-2} & \colhead{2-10}}

\startdata
\multicolumn{4}{l}{From PSD}\\
\hspace{1em} UPS  & ﻿$﻿0.19 \pm 0.01 $ & ﻿$ 0.36 \pm 0.01 $  & ﻿$﻿ 0.404 \pm 0.007 $  \\
\hspace{1em} UFF  &  $﻿﻿0.29_{-0.25}^{+0.12} $ & ﻿$﻿0.46_{-0.10}^{+0.08}  $  & ﻿$0.0_{-0.0}^{+0.1}$ \\
\hspace{1em} Total &  $﻿0.48_{-0.24}^{+0.11}$  & $ 0.83_{-0.09}^{+0.09} $ & $﻿0.404_{-0.007}^{+0.1}$ \\
\multicolumn{4}{l}{From energy spectrum} \\
\hspace{1em} UPS  & ﻿$﻿0.06 \pm 0.02 $ & ﻿$ 0.6 \pm 0.2 $  & ﻿$﻿ 0.4 \pm 0.1 $  \\
\hspace{1em} UFF  &  $﻿﻿0.3 \pm 0.2 $ & ﻿$﻿0.1 \pm 0.1  $  & ﻿$﻿ 0.01 \pm 0.03 $ \\
\hspace{1em} Total &  $﻿0.4 \pm 0.2$  & $ 0.7 \pm 0.2 $ & $﻿0.4 \pm 0.1$ \\
\multicolumn{4}{l}{From data} \\
\hspace{1em} Total &  $0.41 \pm 0.20$  & $ 0.72 \pm 0.19$ & $0.53\pm 0.23$ \\
\enddata
\tablenotetext{}{Counting rates are in unit of counts ${\rm s^{-1}~cm^{-2}~deg^{-2}}$ in terms of the \textit{Suzaku} PI energy band.}
\end{deluxetable}

\section{Discussion}
\label{section:4}

\subsection{Diffuse fraction in 0.5-2 keV; comparison with the energy spectra}
\label{sec:energy_spec}
We have derived the spatial PSD of the X-ray background in a length scale around 50$\sim$1000 
arc second for three different energy bands. The PSD in 2-10 keV can be well described with a
model representing the unresolved point sources (UPS).  This result is consistent with the 
previous results that about 90 \% of the 2-10 keV X-ray background was resolved into point sources and that the rest of the emission is likely from fainter sources.   
On the other hand, the PSD of 0.5-2 keV band requires an additional PSD model representing the uniform flat field (UFF) emission.   
We estimated the counting rate of the two PSD components and the results inferred that $56_{-7}^{+5}$\% of the total count rate 
in this energy band is associated to the diffuse flat-field emission.  
For 2-10 keV, we obtained an upper limit of 23\% for the diffuse emission.   
For the lowest energy band, 0.2-0.5 keV, the diffuse emission fraction of the total background is $60_{-42}^{+9}$\%.

In this subsection, we will compare those results with the energy spectra obtained with the same \textit{Suzaku} observation.  

We first derived energy spectra of 84 images described in subsection \ref{section:2.1},
and fitted the 84 XIS-1 energy spectra in 0.2-10 keV separately. 
With respect to the spectral model, we followed the same strategy introduced in \citet{Yoshino2009}.  
They fitted the energy spectrum of the X-ray background in $\sim 0.5$ to 10 keV with
a spectral model consisting of two thin-thermal emission components and a double broken power-law component 
(e.g \citet{Yoshino2009}). One of the thin-thermal components is considered to represent a sum of the solar-wind induced charge exchange (SWCX) emission and the emission from the local bubble,  
modeled by APEC in XSPEC, which is not subject to the Galactic absorption. The other thin-thermal component is considered to arise from 
the hot gas in the halo of our Galaxy, thus it is subject to the Galactic absorption.  
The double broken power-law component is considered to represent the emission from the extragalactic faint point sources 
and is subject to the Galactic absorption. This component is a sum of the two broken power-law functions.  
The two power-law photon indices are fixed at 1.96 and 1.54, respectively, below 1.2 keV,
while both indices are fixed at 1.4 above 1.2 keV. The normalization factor of the power-law function 
with an index of 1.54 below 1.2 keV is fixed at 5.7 $\rm photon\ s^{-1}\ cm^{-2}\ keV^{-1}\ Sr^{-1}$ at 1 keV, 
and only the normalization factor of the other power-law component is set free for spectrum fitting.  
We fixed the absorption column density to the value estimated from 21 cm observations\citep{HI4PI2016}.   
The free parameters of the spectrum fitting were five: three normalization factors and the temperatures of the absorbed and the unabsorbed thin-thermal components.
The resultant reduced $\chi^2$ values were in the range of 0.95 to 3.99 for the $d.o.f$ of 99 to 640,
and were primarily smaller than 2 for more than 95\% of the spectrum fitting results.
We then estimated the counting rates for each spectral model components that are convolved with 
the response functions of the instrument in all three energy bands, 
and calculated the averages and the standard deviations. 

The counting rate of the double broken power-law component is considered to correspond to 
that of the UPS component of the PSD.  On the other hand, the sum of the counting rates of 
the two thin-thermal components correspond to that of the UFF. As shown in \Tab \ref{tab:count_rates}, 
total counting rates from the energy spectra and the PSD are consistent with each other.  
The fraction of the diffuse emission to the total was estimated to be $82 \pm 10$\%, $15 \pm 10$\%, and $1 \pm 5$\%
for 0.2-0.5, 0.5-2, and 2-10 keV, respectively.
The diffuse emission fraction for 0.5-2 keV band, 15\%, is significantly smaller than the value obtained from the PSD fit.

\citet{McCammon2002} tried to constrain the fraction of thermal
emission combining line emission intensities of the X-ray background observed by the X-ray microcalorimeter
array onboard the XQC sounding rocket and the \textit{ROSAT} R4 band intensity. They estimated
that OVII and OVIII emission lines accounts for 32\% of the X-ray background in \textit{ROSAT} R4 band.
Adding possible thermal continuum emission associated with the Oxygen lines, they estimated
that at least 42\% of the total rate is of thermal origin which is spatially diffuse. 
They also claimed that 38\% of the R4 band is accounted for by the AGNs, leaving 20\% for 
possible diffuse extragalactic contribution.  Thus the diffuse fraction can be as large as 62\%.

In order to compare our result with \citet{McCammon2002}, we need to convert the counting rate of the \textit{ROSAT} R4 band to that of the  \textit{Suzaku} 0.5-2 keV band.
Assuming the energy spectrum described above, i.e. double broken power-law functions for the unresolved faint point sources, 
and a sum of absorbed and unabsorbed thin thermal emission for the diffuse emission. 
The spectral prameters are fixed to the average value from the aforementioned \textit{Suzaku} spectral analysis. 
We found that the diffuse fraction of 42\% and 62\% in the ROSAT R4 band are respectively converted to 16\% and 30\% for the \textit{Suzaku} 0.5-2 keV band.
The former value, i.e. the fraction of thermal emission, is consistent with that from the spectral fitting of \textit{Suzaku} spectra
($15\pm10$\%).

We conclude that the present analysis of the spatial PSD measures a larger diffuse fraction compared to 
the values estimated from the energy spectra of both XQC and \textit{Suzaku}.
Systematic effects in both or either of the PSD and energy-spectrum analyses may explain this discrepancy. 

In the energy spectrum analysis, the intensities of the \ion{O}{7} and \ion{O}{8} lines are determined directly from the energy spectrum.  
Those lines are likely to arise from three different origins, the SWCX, the local bubble, and the hot gas in the halo of our Galaxy.  
The fractions of three origins in line intensities cannot be determined from spectral analysis alone which require some assumptions \citep[e.g.][]{Yoshino2009}. 
The continuum intensity of thermal emission is very sensitive to the plasma temperature. For example at $kT \sim 0.2$ keV, the \ion{O}{7} emissivity of a CIE plasma decreases rapidly with increasing temperature, as a function of  $T^{-3.8}$, while the continuum decreases only slowly.  
Since the \ion{O}{8} to \ion{O}{7} intensity ratio mainly constrain the plasma temperature, the continuum intensity is very dependent on the the component fraction of the lines in the model we assume.
Furthermore, we have almost no constraint on the metal abundance of the plasma. 
In the \textit{Suzaku} spectral fit, the double broken-power-law model also contribute a substantial fraction of the background in 0.5-2 keV where thermal emission start to dominate. 
As a result, the intensity of the continuum of the thermal emission must be strongly coupled with the double broken-power-law model.
Changing the second power-law index of the double broken-power-law model from 1.96 to 1.0 and set both the abundance and the temperature of the the $\sim 0.2$ keV APEC free in the spectral fitting,  we found the diffuse fraction was boosted to $27 \pm 14$\%.  

The other uncertainty of spectral analysis is possible existence of extra-galactic diffuse emission, which is not included in the spectral analysis.  
If exists, it will not only increase the diffuse fraction but also modify, likely flatten, the spectral shape of emission underneath the Galactic emission.  

The PSD analysis also contains systematic errors. We adopted a single value of $S_{\rm max}$ in the model, 
while we averaged PSDs of different images which should have different values of $S_{\rm max}$.  
This is likely to introduce a systematic error in the counting rate of the UPS model.  
Since the sum of the counting rates of the UPS and UFF models is consistent with that of the 
energy spectrum within the error, the total counts needs be kept within the statistical uncertainty 
even when the counting rate of the UPS model is modified because of the systematic error.  
Then the lower limit of the diffuse fraction is estimated to be 40\%, using the maximum 
total count rate from the energy spectrum and the minimum UFF counting rate from the PSD analysis. 
This is marginally consistent with the upper bound of the  diffuse fraction, 41\%, 
from the spectral analysis with modified power-law index above.

In summary, by taking possible systematic effects of both PSD and energy spectrum analyses into account, 
we consider that we can compromise the discrepancy and we conjecture that the value is around 40\%.

\subsection{Diffuse fraction in 0.2-0.5 keV}
\label{sec:diffuse_0.2-0.5}

We are also able to compare the diffuse fraction from the spatial PSD with that from the spectral analysis in 0.2-0.5 keV.  
From the model fits of the \textit{Suzaku} spectra we estimate that the diffuse fraction is $82 \pm 10$\% for the soft energy band of 0.2-0.5 keV. 
This value is consistent with the value from the present PSD analysis within the statistical errors.  
Our lower limit of the faint point-source contribution is 30\%. It suggests that the extragalactic 
point sources with steep power-law index of $\sim 2$ is still contributing to the X-ray background in this energy band 
in spite of the large Galactic absorption.

\section{Summary and conclusion}
\label{section:5}

We studied the spatial power spectrum densities (PSDs) for the blank X-ray skies observed with \textit{Suzaku} in three energy bands separately utilizing the modified $\Delta$-variance method.
We constructed two types of model PSD functions.  One model represents the PSD of the 
unresolved point sources (UPS) which follow the $\log N -\log S$ relations estimated 
from the \textit{Chandra} and \textit{XMM-Newton} deep field observations.
The other model is to account for a uniform flat field (UFF), which we consider to represent 
the truly diffuse emission. The observed PSDs were fitted with a model function consisting of a sum of the two model components.  
We can estimate the counting rates of the UPS and UFF components that best fit the observation.
The main conclusions of the fit results are summarized as follows:
\begin{enumerate}[(1)]
\item The observed PSDs can be fitted well with the model function.
\item In the lower two energy bands, 0.2-0.5 and 0.5-2 keV, both the UPS and UFF components
are necessary to reproduce the observed PSD. However, for 2-10 keV, only an upper limit was obtained for the UFF. 
\item The sums of the UPS and UFF counting rates are consistent with the counting rates actually observed by the X-ray detector 
in all three energy bands.
\item The fractions of the UFF counting rates to the total, namely the diffuse emission fraction of the unresolved X-ray 
background, are estimated to be $60_{-42}^{+9}$\%, $56_{-7}^{+5}$\%, and $ <23$\% in 0.2-0.5 keV, 0.5-2, 2-10 keV, respectively.  
\item The diffuse emission fraction can also be estimated from the energy spectra, where the emission lines 
and their associated continuum counterpart assuming a thermal plasma in the collisional ionization equilibrium state 
are considered to represent for the truly diffuse component.  
For 0.5-2 keV, the diffuse fraction estimated by the present PSD analysis is significantly larger than 
that from the energy spectra, whereas for 0.2-0.5 keV, they are consistent with each other within the large statistical errors.
\item Systematic effects in the energy spectrum model and in the PSD analysis was investigated as the cause of the discrepancy of 0.5-2 keV band.  
We conjecture that these effects can reconcile the discrepancy.

\end{enumerate}

In conclusion, we have demonstrated that even with the limited spatial resolution of 
the \textit{Suzaku} X-ray telescope, the spatial power spectrum is a powerful tool to constrain 
the origins of the X-ray emission.   
Future observations with wider field of view and/or better spatial resolution will 
elucidate the origins in more details and more conclusively.

\acknowledgments
{This work is partially supported by Grants-in-Aid for Scientific Research (KAKENHI) from the JSPS 
(No. 18H01260) and the publication is supported by Leading Initiative for Excellent Young Researchers (LEADER) Program from MEXT of Japan.
}

\bibliographystyle{aasjournal}
\bibliography{refs}

\end{document}